\documentclass[11pt]{article}
\usepackage{amssymb}
\usepackage{amsmath}
\usepackage{amscd}
\usepackage{latexsym}
\usepackage{bbm}

\catcode`@=11
\renewcommand{\section}{\@startsection{section}{1}{0pt}{\medskipamount}
{\medskipamount}{\large\bf}}
\numberwithin{equation}{section}
\catcode`@=12

\def\a{\alpha}
\def\b{\beta}

\def\z{\zeta}

\def\th{\theta}

\def\ka{\kappa}

\def\s{\sigma}

\newcommand{\C}{\mathbb C}
\newcommand{\R}{\mathbb R}
\newcommand{\Z}{\mathbb Z}
\newcommand{\NN}{{\mathbbm{N}}}
\newcommand{\unity}{\mathbbm{1}}
\newcommand{\Hcal}{{\mathcal H}}
\newcommand{\adag}{a^{\dagger}}
\newcommand{\bp}{\mbox{\boldmath$\phi$}}
\newcommand{\bn}{\mbox{\boldmath$n$}}

\def\im{\textrm{i}}
\def\ep{\textrm{e}}
\def\diff{\textrm{d}}

\def\Tr{\textrm{Tr}}
\def\tU{\textrm{U}}

\def\sfrac#1#2{{\textstyle\frac{#1}{#2}}}
\def\pa{\partial}
\def\>{\rangle}
\def\<{\langle}
\def\+{\dagger}
\def\={\ =\ }
\def\zb{{\bar{z}}}

\begin{document}

\begin{titlepage}
\setcounter{page}{0}
\begin{flushright}
ITP--UH--07/09
\end{flushright}

\vskip 2.0cm

\begin{center}

{\Large\bf  Noncommutative Baby Skyrmions}

\vspace{12mm}

{\large Theodora Ioannidou$\,{}^*$ \ and \ \large Olaf Lechtenfeld$\,{}^+$
}
\\[8mm]
\noindent ${}^*${\em Department of Mathematics, Physics and
Computational Sciences, Faculty of Engineering,\\
Aristotle University of Thessaloniki, Thessaloniki 54124, Greece }\\
{Email: ti3@auth.gr}
\\[8mm]
\noindent ${}^+${\em Institut f\"ur Theoretische Physik,
Leibniz Universit\"at Hannover \\
Appelstra\ss{}e 2, 30167 Hannover, Germany }\\
{Email: lechtenf@itp.uni-hannover.de}
\\[8mm]

\vspace{12mm}

\begin{abstract}

\noindent
We subject the baby Skyrme model to a Moyal deformation, for unitary or 
Grassmannian target spaces and without a potential term. In the abelian case, 
the radial BPS configurations of the ordinary noncommutative sigma model also
solve the baby Skyrme equation of motion. This gives a class of exact analytic
noncommutative baby Skyrmions, which have a singular commutative limit but are
stable against scaling due to the noncommutativity. We compute their energies, 
investigate their stability and determine the asymptotic two-Skyrmion 
interaction.
\end{abstract}

\end{center}
\end{titlepage}

\section{Introduction}

\noindent

It is known that the two-dimensional $\C P^1$~$\sigma$-model~\cite{1} possesses
metastable states which when perturbed may shrink or spread out
due to the conformal (scale) invariance of the model~\cite{2,3,4}.
This implies that the metastable states can be of any size, and so
a four-derivative term, the so-called Skyrme term,
needs to be added for breaking the scale invariance of the model~\cite{PZMKT}.
However, the resulting energy functional has no minima, and
a further, so-called potential (or mass) term is needed to stabilize 
the size of the corresponding solutions. The ensueing model is known 
as the {\it baby Skyrme model}, and it admits stable field configurations 
of solitonic nature called {\it baby Skyrmions}, which can be determined
numerically~\cite{5}. As the extra terms contribute to the masses of the 
solitons, the Skyrmion mass is strictly larger than the Bogomol'nyi bound 
given by the topological charge (Skyrmion number), and the two-Skyrmion
configuration becomes stable showing the existence of bound states~\cite{5}.

In the $\C P^1$~baby Skyrme model, the target manifold~$S^2$ is parametrized 
by a three-dimensional isovector scalar~$\bp$ subject to the constraint 
$|\bp|^2=1$. Its Lagrangian density is of the form
\begin{equation} \label{bs}
{\cal L}\=\sfrac{1}{2}\pa_\mu \bp\, \pa^\mu \bp\ -\
\sfrac{\ka^2}{4}(\pa_\mu\bp\times \pa_\nu \bp)(\pa^\mu\bp\times \pa^\nu \bp)
\ -\ V(\bp) \ ,
\end{equation}
where the field $\bp$ is a map from the three-dimensional Minkowski 
space~$\R^{1,2}$ with the metric 
$(\eta_{\mu\nu})=\textrm{diag}({+}1,{-}1,{-}1)$ to the two-sphere~$S^2$
of unit radius. 
The first term in~(\ref{bs}) is the familiar $\C P^1$~sigma model,
the second term is the two-dimensional analogue of the Skyrme term
and carries a coupling~$\ka$ of the dimension of length, and the last term 
is the potential, for which different proposals have been made.
For \ $V\sim1-(\bn\bp)^2$ \ (the so called new baby Skyrme model)
approximate baby skyrmions were obtained (analytically) by exploring
its topological properties~\cite{IKZ}.
Finiteness of the energy requires the field to approach a zero of the
potential (the `vacuum'~$\bn$) at spatial infinity, allowing one to compactify
the static base space $\R^2$ to $S^2$ and to consider $\bp$ as a map 
$S^2\to S^2$. This gives rise to the homotopy invariant
\begin{equation}
\textrm{deg}[\bp]\=\sfrac{1}{4\pi}{\textstyle\int}\diff{x}\,\diff{y}\ 
\bp \cdot (\pa_x\bp\times \pa_y\bp) \ \in\Z\ ,
\end{equation}
also known as the topological charge or the Skyrmion number, 
which is conserved.

The baby Skyrme model is a useful laboratory for studying soliton physics.
It is the $2{+}1$ dimensional analog of a model which 
describes the low-energy chiral dynamics of Quantum Chromodynamics~\cite{ANW}, 
the usual Skyrme model~\cite{S}. This model has direct applications in 
condensed matter physics~\cite{Mac}, where baby Skyrmions give an effective 
description in quantum Hall systems. In such systems, the dynamics are governed
by the spin stiffness term, the Coulomb interaction and the Zeeman interaction.
In particular, its kinetic energy corresponds to the spin stiffness term, 
and the potential (or mass) term corresponds to the Zeeman interaction, 
the correspondence being exact for the static sector. The Skyrme term is 
analogous to the Coulomb term. All terms are needed to prevent the collapse 
of topological configurations which yield to Skyrmion solutions.

In this situation, 
a {\it noncommutative deformation\/}~(for reviews see~\cite{nc}) 
may serve as a substitute for the potential term (or Zeeman interaction), 
because it introduces a new length scale into the theory, which also 
stabilizes solitons against collapse or spreading.
We expect this to give rise to a new class of baby Skyrmions.
Indeed, it is known that Moyal-deformed field theories have a much
richer soliton spectrum than their commutative counterparts
(see, e.g.,~\cite{lepo01,sendai} and references therein).

Furthermore, it is not easy to access the quantum fluctuations in the Skyrme 
and baby Skyrme models, since the field theories are perturbatively 
non-renormalizable and thus, existing treatments are semiclassical 
(quantizing only the collective degrees of freedom of the soliton). 
Full quantization of the theory requires a cutoff which can be attained by its 
lattice version. Here, a noncommutative deformation may again be of help,
since it introduces a regulating parameter into the quantum theory.
Quite generally, the noncommutative version of a theory may improve its 
renormalizability properties at short distances and may even render it finite. 
The two above applications of noncommutativity are our main motivation
for Moyal-deforming the baby Skyrme model.

In this Letter, we present a noncommutative baby Skyrme model,\footnote{
Different aspects of Moyal-deforming a Skyrme model have appeared 
in~\cite{mieck}.}
without potential term, for group- or Grassmannian-valued targets, 
and explicitly obtain a class of exact analytic solitonic solutions, 
which have no analogues in the commutative theory.
This surprising feat succeeds because certain BPS configurations of the
Moyal-deformed ordinary sigma model extremize the Skyrme part of the energy
as well. We compute their static energy, discuss their stability and
evaluate the two-Skyrmion interaction potential at large distances.


\section{The baby Skyrme model}
\noindent
The $\C P^1$~sigma model is the paradigm of a Grassmannian sigma model,
since the target manifold can be written as
\begin{equation}
\C P^1\ \simeq\ S^2\ \simeq\ \frac{\textrm{SU}(2)}{\tU(1)}\ \simeq\ 
\frac{\tU(2)}{\tU(1)\times\tU(1)}\ \simeq\ \textrm{Gr}(2,1)\ ,
\end{equation}
with the definition 
\begin{equation}
\textrm{Gr}(n,k)\ :=\ \frac{\tU(n)}{\tU(k)\times\tU(n{-}k)}\ \simeq\
\frac{\tU(\C^n)}{\tU(\textrm{im}P)\times\tU(\textrm{ker}P)}
\qquad\textrm{for a projector $P$ of rank $k$} \ .
\end{equation}
A general group-valued or Grassmannian-valued baby Skyrme model then 
features fields
\begin{equation}
g:\ \R^{1,2}\ \to\ \tU(n)\ \ \textrm{or}\ \ \textrm{Gr}(n,k)
\qquad\textrm{via}\qquad(x^\mu)\equiv(t,x^i)\equiv(t,x,y)\ \mapsto\ g(t,x,y)\ ,
\end{equation}
which enter as variables in the action (without potential term)
\begin{equation}
S\=\int\!\!\diff^{1+2}x\ 
\Bigl\{\sfrac12\,\eta^{\mu\nu}\pa_\mu g^\+\,\pa_\nu g\ +\ \sfrac{\ka^2}{4}\,
[\,g^\+\pa_\mu g\,,\,g^\+\pa_\nu g\,]\ [\,g^\+\pa^\mu g\,,\,g^\+\pa^\nu g\,]
\Bigr\}\ .
\end{equation}
Classical solutions are obtained by solving the equation of motion
\begin{equation} 
\pa^\mu j_\mu\=0 \qquad\textrm{for}\qquad
j_\mu\= g^\+\pa_\mu g\ +\ \ka^2\,
\bigl[\,g^\+\pa^\nu g\,,\,[\,g^\+\pa_\mu g\,,\,g^\+\pa_\nu g\,]\,\bigr]\ .
\end{equation}

Let us concentrate on static solutions, $\pa_t g\equiv0$,
which are found by extremizing the energy
\begin{equation} \label{energy}
E\=\int\!\!\diff^2x\ \Bigl\{\sfrac12\,\pa_ig^\+\,\pa_ig\ -\ \sfrac{\ka^2}{4}\,
[\,g^\+\pa_i g\,,\,g^\+\pa_j g\,]\ [\,g^\+\pa_i g\,,\,g^\+\pa_j g\,] \Bigr\}\ .
\end{equation}
For Grassmannian models, this simplifies since Gr$(n,k)$ is embedded
in U($n$) via the constraint
\begin{equation} \label{Gr}
g^2=\unity_n \qquad\Leftrightarrow\qquad g^\+=g \qquad\Leftrightarrow\qquad
g=\unity_n-2P \quad\textrm{with}\quad P^\+=P=P^2\ ,
\end{equation}
and so their energy becomes
\begin{equation} \label{Grenergy}
E_{\textrm{Gr}}\=
\int\!\!\diff^2x\ \Bigl\{2\,P_i\,P_i\ -\ 4\ka^2\,[P_i,P_j]\,[P_i,P_j]\Bigr\}\ ,
\end{equation}
where the standard notation \ $\pa_iP=P_i$~, $\pa_i\pa_jP=P_{ij}$ \ etc.~was
introduced.
We are looking for extrema of the energy~(\ref{energy}) which are located
inside some Grassmannian. Putting $\delta E=0$ and employing~(\ref{Gr}),
in particular \ 
$g^\+\pa g=-2[\pa P,P]$ \ and \ $\pa_i(g^\+\pa_i g)=-2[P_{ii},P]$~, 
one arrives at
\begin{equation} \label{eom}
[\,P_{ii}\,,\,P\,]\ +\ 4\ka^2\,F[P] \= 0 \qquad\textrm{with} 
\end{equation}
\begin{equation}
F[P]\= 2P_{ij}[P_i,P]P_j\ -\ \pa_i(P_jP_j)[P_i,P]\
+\ P_j[P_{ii},P]P_j\ -\ P_jP_j[P_{ii},P]\ -\ \textrm{h.c.}\ .
\end{equation}
Solutions to (\ref{eom}) extremize the energy~(\ref{Grenergy}) of the Gr$(n,k)$
model as well as (stronger) the energy~(\ref{energy}) of the U($n$)~model.
{}From now on we pass to complex coordinates
\begin{equation}
\biggl. \begin{aligned} z&\=x+\im y \\ \zb&\=x-\im y \end{aligned}\quad\biggr\}
\qquad\Longrightarrow\qquad
\biggl\{ \begin{aligned} \pa_x &\= \pa_z+\pa_\zb \\ -\im\pa_y &\= \pa_z-\pa_\zb
\end{aligned}\biggr. \quad.
\end{equation}

At $\ka=0$ we connect with the ordinary sigma model.
Grassmannian-valued extrema of its energy are provided by the 
well known BPS projectors, defined through
\begin{equation} \label{BPS}
0 \= (\unity_n{-}P)\,P_\zb \= P_\zb\,P \qquad\Longleftrightarrow\qquad
0 \= P_z\,(\unity_n{-}P) \= P\,P_z\ .
\end{equation}
These relations (together with $P^2=P$) imply various useful identities, 
such as \ $[P_z,P_\zb]=P_{z\zb}$ \ and
\begin{equation}
0 \= (\unity_n{-}P)\,P_{\zb\zb} \= P_{\zb\zb}\,P
  \= P_{zz}\,(\unity_n{-}P) \= P\,P_{zz}
  \= P_z\,P_z \= P_\zb\,P_\zb \= [P_{z\zb},P]\ .
\end{equation}
We now turn $\ka$ back on and compute the failure of the BPS projectors to
extremize the baby Skyrme energy:
\begin{equation} \label{failure}
\sfrac18\,F[P\ \textrm{subject to (\ref{BPS})}] \=
K_z P_\zb - K_\zb P_z - P_z K_\zb + P_\zb K_z \=
P_\zb\,P_{zz}\,P_\zb\ -\ P_z\,P_{\zb\zb}\,P_z
\end{equation}
with the definition \ $K\equiv\sfrac14P_iP_i=\sfrac12(P_z P_\zb+P_\zb P_z)$~.

To get a feeling, we evaluate this expression in the $\C P^1$ model for the 
(rank-one) BPS projectors, which are based on holomorphic functions~$f$,
\begin{equation}
\begin{aligned}
& P\=\sfrac{1}{1+f\bar{f}}\,
\Bigl(\begin{smallmatrix}1&\ \bar{f}\\[4pt]f&\ f\bar{f}\end{smallmatrix}\Bigr)
\qquad\Longrightarrow\qquad
2K\=\sfrac{f'\bar{f}'}{(1+f\bar{f})^2}\,\unity_2
\qquad\Longrightarrow\qquad \\[8pt]
& \sfrac18\,F\=\sfrac{1}{(1+f\bar{f})^4}\,
\biggl(\begin{smallmatrix} \bar{f}{f'}^2\bar{f}''-f\bar{f}^{\prime2}f'' \ & \
\bar{f}^2{f'}^2\bar{f}''+\bar{f}^{\prime2}f''-2\bar{f}{f'}^2\bar{f}^{\prime2} 
\\[6pt]
-f^2\bar{f}^{\prime2}f''-{f'}^2\bar{f}''+2f{f'}^2\bar{f}^{\prime2} \ & \
f\bar{f}^{\prime2}f''-\bar{f}{f'}^2\bar{f}'' \end{smallmatrix} \biggr)\ .
\end{aligned}
\end{equation}
This vanishes only for constant~$f$.
Even in the simplest case, $f=z$, one finds \ $\sfrac18F=\sfrac2{(1+z\zb)^4}
\bigl(\begin{smallmatrix}0&-\zb\\z&\phantom{-}0\end{smallmatrix}\bigr)$.
We conclude that the sigma-model BPS~solitons never obey the baby Skyrme
equation of motion.


\section{Moyal deformation and abelian model}
\noindent
A Moyal deformation of Euclidean $\R^2$ with coordinates $(x,y)$
is achieved by replacing the ordinary pointwise product of smooth functions
on it with the noncommutative but associative Moyal star product. 
The latter is characterized by a constant positive real parameter~$\th$ which
prominently appears in the star commutation relation between the coordinates,
\begin{equation} \label{nccoord}
x\star y - y\star x \ \equiv\ [\,x\,,\,y\,]_\star \= \im\,\th 
\qquad\Longrightarrow\qquad [\,z\,,\,\bar{z}\,]_\star \= 2\th\ .
\end{equation}
It is convenient to work with the dimensionless coordinates
\begin{equation}
a \= \tfrac{z}{\sqrt{2\th}} \quad\textrm{and}\quad 
\adag \= \tfrac{\bar{z}}{\sqrt{2\th}}
\qquad\Longrightarrow\qquad [\,a\,,\,\adag\,]_\star \= 1 \ . 
\end{equation}
For a concise treatment of the Moyal star product see~\cite{nc}.

A different realization of this Heisenberg algebra promotes the coordinates
(and thus all their functions) to noncommuting operators acting on an
auxiliary Fock space~$\Hcal$ but keeps the ordinary operator product.
The Fock space is a Hilbert space with orthonormal basis states
\begin{equation} \label{oscbasis}
\begin{aligned}
{}& \qquad\qquad\qquad\quad |m\>\=\sfrac{1}{\sqrt{m!}}\,(\adag)^m\,|0\>
\qquad\text{for}\quad m\in\NN_0 \quad\text{and}\quad a|0\>=0 \ ,\\[4pt]
{}& a\,|m\> \= \sqrt{m}\,|m{-}1\> \ ,\qquad
\adag\,|m\> \= \sqrt{m{+}1}\,|m{+}1\> \ ,\qquad
N\,|m\> \ := \adag a\,|m\> \= m\,|m\> \ ,
\end{aligned}
\end{equation}
therewith characterizing $a$ and $\adag$ as standard annihilation 
and creation operators. The star-product and operator formulations are
tightly connected through the Moyal-Weyl map:
Coordinate derivatives correspond to commutators with coordinate operators,
\begin{equation} \label{derivs}
\sqrt{2\th}\,\pa_z \ \leftrightarrow\ -\textrm{ad}(\adag)\quad,\qquad
\sqrt{2\th}\,\pa_{\bar z} \ \leftrightarrow\ \textrm{ad}(a) \quad,
\end{equation}
and the integral over the noncommutative plane reads
\begin{equation}
\int\!\diff^2x\;f_{\star}(x) \= 2\pi\,\th\;\Tr_{\Hcal}\,f_{\mathrm{op}} \ ,
\end{equation}
where the function $f_{\star}$ corresponds to the operator~$f_{\mathrm{op}}$
via the Moyal-Weyl map and the trace is over the Fock space~$\Hcal$.
We shall work with the operator formalism but refrain from introducing
special notation indicating operators, so all objects are operator-valued
if not said otherwise. The time coordinate~$t$ of the full baby Skyrme model
remains commutative. Hence, we trade the spatial dependence of our fields
with operator valuedness (in~$\Hcal$), and thus work with maps from the
time interval into an enlarged target space, namely \ 
$\tU(\C^n\otimes\Hcal)= \tU(\Hcal\oplus\ldots\oplus\Hcal)$ \ or some
Grassmannian subspace thereof.

Since the noncommutative target space is much bigger than the original one,
new possibilities for BPS projectors arise. In fact, the classical solutions
to the deformed theory come in two types:
Firstly, {\it nonabelian solutions\/} are continuously (in~$\th$) connected to
their commutative counterparts (tensored with $\unity_\Hcal$) and represent 
smooth deformations of it. Secondly, {\it abelian solutions\/} become
singular at $\th\to0$ and are genuinely noncommutative. In the BPS case,
the simplest abelian projectors are of finite rank or co-rank in one copy
of~$\Hcal$. Since novel features can be expected only in the abelian case,
we focus on it from now on and choose $n{=}1$, i.e.~the Moyal-deformed
U(1)~baby Skyrme model. Clearly, this theory permits abelian solutions only,
since its commutative limit is free. However, it still contains an infinity
of Grassmannian submodels corresponding to \ 
$\textrm{Gr}(P)=\frac{\tU(\Hcal)}{\tU(\textrm{im}P)\times\tU(\textrm{ker}P)}$
\ for some hermitian projector~$P$, preferably of finite rank or co-rank~$k$.

The Moyal deformation introduces the dimensionful parameter~$\th$ into the
theory, which invalidates Derrick's argument: scaling of spatial coordinates
now relates theories with different strengths of noncommutativity.
Therefore, classical solutions at a fixed value of~$\th$ are safe against
shrinking or spreading. 


\section{Exact noncommutative baby Skyrmions}
\noindent
The equations of section~2 carry over to the deformed abelian baby Skyrme model
(with replacing $\unity_n$ by $\unity_\Hcal$), since on a formal level its
noncommutativity resembles the non-abelianness in the standard U($n$)~model. 
Hence, the failure of a standard noncommutative U(1)~sigma-model BPS~solution,
$g=\unity{-}2P$ obeying~(\ref{BPS}), to also fulfil the baby Skyrme equation 
of motion, is again measured by~(\ref{failure}).
In our Moyal-deformed theory, this expression may vanish, and surprisingly does
so if the projector is a function of the number operator $N{=}\,\adag a$ only!
In the star-product picture, this corresponds to functions only of the radial
variable $r{=}\sqrt{z\zb}$, and so they are called {\it radial projectors\/}. 
It is obvious that $F[P]$ in~(\ref{failure}) vanishes for $P=P(r)$, but
in the commutative theory only trivial projectors can be radial. In the
Fock-space basis~(\ref{oscbasis}), radial projectors are simply diagonal. 

Indeed, it is not hard to check explicitly that the BPS projector
\begin{equation} \label{Pk}
P^{(k)} \ :=\ \sum_{n=0}^{k-1} |n\>\<n| \qquad\textrm{obeys}\qquad
P^{(k)}_\zb P^{(k)}_{zz} P^{(k)}_\zb \=0\=
P^{(k)}_z P^{(k)}_{\zb\zb} P^{(k)}_z
\end{equation}
as well as \ $[P^{(k)}_{z\zb},P^{(k)}]=0$~, 
in the sense of (\ref{derivs}).
Hence, $F[P^{(k)}]=0$, and the noncommutative baby Skyrme equation of motion
is satisfied.
In addition, due to the translation invariance of the model, the translates
\begin{equation}
P^{(k,\a)}\ :=\ \ep^{\a\adag-\bar\a a}\,P^{(k)}\,\ep^{-\a\adag+\bar\a a}
\qquad\textrm{for}\quad \a\in\C \quad\textrm{and}\quad k\in\NN
\end{equation}
also do the job. It is noteworthy that the role of $\textrm{deg}[\bp]$ for
the topological charge has been taken by the rank~$k$ of the projector,
which also defines a Grassmannian submanifold.
Thus, for each value of~$k$ we have found a $\C$-family of exact noncommutative
U(1)-valued baby Skyrmions, which are of course also solitons in the 
Grassmannian submodel. Most basic is the $k{=}1$ family
\begin{equation}
P^{(1,\a)}\=\ep^{-\bar\a\a}\,\ep^{\a\adag}|0\>\<0|\,\ep^{\bar\a a}
\ =:\ |\a\>\<\a| \qquad\textrm{with}\quad a|\a\>=\a|\a\>\ ,
\end{equation}
which consists of the coherent-state projectors obtained by translating
the ground-state projector \ $P^{(1)}=|0\>\<0|$.\footnote{
One may wonder how this can be given as a function of~$N$. 
One possiblity is \ $|0\>\<0|=\frac{\sin\pi N}{\pi N}~$.}
The corresponding function (under the Moyal-Weyl map) is just a Gaussian
centered at~$\a$ in the Moyal plane,
\begin{equation}
P^{(1,\a)}_\star(z,\zb) \= 2\,\ep^{-|z-\a|^2/\th}\ ,
\end{equation}
and the singular $\th\to0$ limit becomes apparent.

Let us take a look at the energy of these configurations.
The Grassmannian energy functional~(\ref{Grenergy}) reads
\begin{equation}
E_{\textrm{Gr}}[P] \= 16\pi\th\,\Tr_\Hcal \bigl\{
P_z\,P_\zb\ +\ 4\ka^2\,[P_z,P_\zb]^2 \bigr\}
\end{equation}
which for BPS projectors, due to \ $[P_z,P_\zb]=P_{z\zb}$~, simplifies to
\begin{equation}
E_{\textrm{BPS}}[P] \= 16\pi\th\,\Tr_\Hcal \bigl\{ 
P_z\,P_\zb\ +\ 4\ka^2\,P_{z\zb}^2 \bigr\} \=
8\pi\,\Tr_\Hcal \bigl\{ -[\adag,P]\,[a,P]\ +\ 
\sfrac{2\ka^2}{\th}\,[\adag,[a,P]\,]^2 \bigr\} \ .
\end{equation}
It is straightforward to evaluate this on the rank-$k$ diagonal projector
of~(\ref{Pk}),
\begin{equation} \label{kenergy}
E[P^{(k)}] \= 8\pi\,\Tr_\Hcal \bigl\{ k\,|k\>\<k|\ +\ 
\sfrac{2\ka^2}{\th}\,k^2\,\bigl(|k\>\<k|+|k{+}1\>\<k{+}1|\bigr)\bigr\}\=
8\pi\,\bigl(k\,+\,\sfrac{4\ka^2}{\th}\,k^2\bigr)\ .
\end{equation}
Due to translation invariance, the same result holds for~$P^{(k,\a)}$.
The energy depends only on the dimensionless parameter $\ka^2/\th$. 
It exceeds the Bogomol'nyi bound of $8\pi k$ by the contribution of 
the Skyrme term, whose $k^2$ dependence signals an instability of
the higher-charge baby Skyrmions against decay into those of charge one.
Interpreting $P^{(k)}$ as describing $k$~charge-one baby Skyrmions
sitting on top of each other, they can lower their energy by passing to a
configuration of near-infinite mutual separation, which is again a
(near-exact) baby Skyrme solution. More general multi-center BPS solitons
do not solve the baby Skyrme equation of motion~(\ref{eom}), since
they are not rotationally symmetric, and thus $F[P]$ does not vanish.


\section{Stability and interactions}
\noindent
Are our noncommutative baby Skyrmions stable?
If this question is asked for the full U(1)~model, the answer is negative
by a standard argument:
Consider a path in $\tU(\Hcal)$ which connects a Grassmannian solution 
to the vacuum,
\begin{equation}
g(s) \= \ep^{\im(\pi-s)P} \= \unity\ -\ (1{+}\ep^{-\im s})P
\qquad\textrm{with}\quad P^\+=P=P^2 \quad\textrm{and}\quad s\in[0,\pi]\ .
\end{equation}
It interpolates between \ $g(0)=\unity{-}2P\in\textrm{Gr}(P)$ \ and \ 
$g(\pi)=\unity$. The energy 
\begin{equation}
\begin{aligned}
E(s) &\= 4\pi\th\,\Tr_\Hcal\bigl\{ \pa_z g^\+ \pa_\zb g\ 
+\ \ka^2( \pa_z g^\+ \pa_\zb g - \pa_\zb g^\+ \pa_z g )^2 \bigr\} \\[4pt]
&\= 4\pi\th\,\bigl\{ (1{+}\ep^{\im s})(1{+}\ep^{-\im s})\,\Tr_\Hcal(P_zP_\zb)\ 
+\ \ka^2(1{+}\ep^{\im s})^2(1{+}\ep^{-\im s})^2\,\Tr_\Hcal[P_z,P_\zb]^2\bigr\}
\\[4pt]
&\= 4\pi\th\,\bigl\{ \sfrac{k}{2\th}\cdot 4\cos^2\sfrac{s}{2}\ 
+\ \sfrac{2k^2}{4\th^2}\cdot 16k^2\cos^4\sfrac{s}{2} \bigr\} \=
8\pi\,\bigl\{ k\cos^2\sfrac{s}{2}\ +\ 
\sfrac{4\ka^2}{\th}\,k^2\cos^4\sfrac{s}{2}\bigr\}
\end{aligned}
\end{equation}
along the path is decreasing monotonically to zero, which renders any
soliton of the U(1)~model unstable. This is not surprising, since the 
topological charge is well defined and conserved only inside the Grassmannian
submanifolds. 

So we should ask about the stability of our noncommutative solitons inside a 
Grassmannian baby Skyrme model. The energy formula~(\ref{kenergy}) shows that 
for rank~$k$ the solution~$P^{(k)}$ will decay into $k$ well-separated copies 
of~$P^{(1)}$, so only the {\it charge-one\/} baby Skyrmion may be 
(and probably is) stable. However unlikely, it is still not excluded that 
it can lower its energy by changing its shape away from being round and 
becoming non-BPS. One could settle this issue by computing the second variation
$\delta^2E$ restricted to Gr($P^{(1)}$), which we have left for future work.

To determine the long-range forces between two noncommutative baby Skyrmions,
we compute the energy of a two-center BPS soliton, because for
large separation this configuration approaches a superposition of two
rank-one BPS solitons, which we have already found to be baby Skyrmions.
In the two-center configuration 
\begin{equation}
P^{(\a,\b)}\=\sfrac{1}{1-|\s|^2}\,\bigl\{
|\a\>\<\a|\ +\ |\b\>\<\b|\ -\ \s|\a\>\<\b|\ -\ \bar\s|\b\>\<\a| \bigr\}
\qquad\textrm{with}\quad
\s=\<\a|\b\>
\end{equation}
the lumps are centered at positions $\a$ and $\b$ in the complex Moyal plane,
and the coherent states $|\a\>$ and $|\b\>$ are normalized to one.
This projector obeys the BPS condition~(\ref{BPS}) hence 
$[P^{(\a,\b)}_{z\zb},P^{(\a,\b)}]=0$ but $F[P^{(\a,\b)}]\neq0$ 
unless $\a{-}\b\to0$ or~$\infty$.
Employing the defining relations \ $(a{-}\a)|\a\>=0$ \ and \ $(a{-}\b)|\b\>=0$
\ as well as \ $\s\bar\s=\ep^{-|\a-\b|^2}$, it is straightforward to compute
\begin{equation}
\begin{aligned}
E[P^{(\a,\b)}] &\=
8\pi\,\Tr_\Hcal \bigl\{ -[\adag,P^{(\a,\b)}]\,[a,P^{(\a,\b)}]\ +\ 
\sfrac{2\ka^2}{\th}\,[\adag,[a,P^{(\a,\b)}]\,]^2 \bigr\} \\[8pt]
&\= 8\pi\,\Bigl\{ 2\ +\ 8\,\sfrac{\ka^2}{\th} 
\Bigl( 1 + \sfrac14\,r^4\,\sinh^{-2}\sfrac{r^2}{2} \Bigr) \Bigr\}
\qquad\textrm{with}\quad r:=|\a{-}\b|\ .
\end{aligned}
\end{equation}
This expression interpolates smoothly between
\begin{equation}
E[P^{(r=0)}] \= 8\pi\,(2 + \sfrac{4\ka^2}{\th}\cdot4) \= E[P^{(2)}]
\qquad\textrm{and}\qquad
E[P^{(r\to\infty)}]\=2\cdot8\pi\,(1 + \sfrac{4\ka^2}{\th})\=2\cdot E[P^{(1)}]
\end{equation}
which again underscores the decay channel \ $P^{(2)}\to P^{(1)}+P^{(1)}$.
For large separation, the interaction potential is exponentially repulsive,
\begin{equation}
V(r)\ \sim\ 64\pi\,\sfrac{\ka^2}{\th}\,r^4\,\ep^{-r^2/2}
\qquad\textrm{for}\quad r\to\infty\ .
\end{equation}

We close with a list of open problems. It would be interesting to 
(a) find other exact abelian noncommutative baby Skyrmions or rule out this
possibility, (b) determine whether $P^{(1)}$ has minimal energy in the 
rank-one Grassmannian (i.e.~is stable), and (c) work out the scattering of 
two such lumps. Another promising task is to deform the {\it full\/} Skyrme 
model (on $\R^{1,3}$) and to construct noncommutative Skyrmions from 
noncommutative instantons~\cite{inst}.


\noindent
{\bf Acknowledgements}\\
\noindent
O.L.~is grateful for hospitality extended to him by the Mathematics Division 
of the Department of Mathematics, Physics and Computational Sciences at
Aristotle University. He is partially supported by DFG grant Le-838/9.


\end{document}